 \documentclass[a4,twocolumn,prb,       
superscriptaddress]{revtex4}
   \usepackage{color}
   \usepackage{epsfig}
   \usepackage{graphics}
   \usepackage{epic}

\begin{document}
\title
{X-ray study of structural domains in the near-surface region of SrTiO$_{3}$-substrates with Y$_{0.6}$Pr$_{0.4}$Ba$_{2}$Cu$_{3}$O$_{7}$ / La$_{2/3}$Ca$_{1/3}$MnO$_{3}$ superlattices grown on top}

\author{J. Hoppler}
\email{justin.hoppler@unifr.ch}
\affiliation{Department of Physics and Fribourg Center for Nanomaterials, University of Fribourg, Chemin du Mus\'ee 3, CH-1700 Fribourg}
\affiliation{Laboratory for Neutron Scattering, ETH Zurich \& Paul Scherrer Institut, CH-5232 Villigen PSI, Switzerland}

\author{J. Stahn}
\affiliation{Laboratory for Neutron Scattering, ETH Zurich \& Paul Scherrer Institut, CH-5232 Villigen PSI, Switzerland}

\author{H. Bouyanfif}
\altaffiliation{Now at: LEMA (Laboratory of electrodynamics of advanced materials) UMR 6157 CNRS-CEA, Universit\'e Fran\c cois Rabelais Tours, Facult\'e des Sciences \& Techniques, Parc de Grandmont, 37200 Tours, France}

\author{V. K. Malik}
\affiliation{Department of Physics and Fribourg Center for Nanomaterials, University of Fribourg, Chemin du Mus\'ee 3, CH-1700 Fribourg}

\author{B.D. Patterson}
\affiliation{Swiss Light Source, Paul Scherrer Institut, CH-5232 Villigen PSI, Switzerland}

\author{P.R. Willmott}
\affiliation{Swiss Light Source, Paul Scherrer Institut, CH-5232 Villigen PSI, Switzerland}

\author{G. Cristiani}
\affiliation{Max-Plank-Institut f\"ur Festk\"orperforschung Stuttgart, D-70569 Stuttgart, Germany}

\author{H.-U. Habermeier}
\affiliation{Max-Plank-Institut f\"ur Festk\"orperforschung Stuttgart, D-70569 Stuttgart, Germany}

\author{C. Bernhard}
\affiliation{Department of Physics and Fribourg Center for Nanomaterials, University of Fribourg, Chemin du Mus\'ee 3, CH-1700 Fribourg}

\date{\today}

\begin{abstract}
We investigated with synchrotron x-ray diffraction and reflectometry the formation of structural domains in the near-surface region of single crystalline SrTiO$_3$ (001) substrates with Y$_{0.6}$Pr$_{0.4}$Ba$_{2}$Cu$_{3}$O$_{7}$ / La$_{2/3}$Ca$_{1/3}$MnO$_{3}$ superlattices grown on top. We find that the antiferrodistortive cubic-to-tetragonal transition, which occurs at $T_\mathrm{STO} = 104$\,K in the bulk and at a considerably higher temperature of at least 120\,K in the surface region of SrTiO$_3$, has only a weak influence on the domain formation. The strongest changes occur instead in the vicinitiy of the tetragonal-to-orthorhombic transition in SrTiO$_3$ around 65\,K, where pronounced surface facets develop that reach deep (at least several micrometers) into the SrTiO$_3$ substrate. These micrometer-sized facets are anisotropic and tilted with respect to one another by up to 0.5$^\circ$ along the shorter direction. Finally, we find that a third structural transition below 30\,K gives rise to significant changes in the spread of the $c$-axis parameters. Overall, our data provide evidence for a strong mutual interaction between the stuctural properties of the SrTiO$_3$ surface and the multilayer grown on top.
\end{abstract}

\maketitle
The growth of epitaxial thin films and heterostructures from perovskite-like materials is becoming an increasingly important issue. The interest in these pervoskites is motivated by the extremely rich spectrum of physical phenomena which they can offer like ferroelectricity, magnetism and high temperature superconductivity \cite{Bak98, Sal01, Cav87}. Their similar lateral lattice parameters allow one to tailor multilayers and heterostructures which combine materials with various kinds of competing interactions and order parameters. Recently, it has been demonstrated that this approach can be used to create artificial materials with designed physical properties and even with novel proximity-induced magnetic or electronic states \cite{Sta05, Cha06, San06, Hol04, Rij05, Hui06, Dag07}. 

The most commonly used substrate material for growing these thin films and multilayers is SrTiO$_3$ (STO) with (001) orientation. This is because large single crystals are readily available and it is possible to control their surface termination by chemical etching \cite{Ohn04}. In this context, it has often not been considered that STO has fairly complex structural properties with a series of structural phase transitions (PT) \cite{Lyt64, Bli05}. Beside the antiferrodistortive cubic-to-tetragonal PT at $T_\mathrm{STO}^{I} = 104\,$K there are at least two more PTs. The second PT at $T^{II}_\mathrm{STO} = 65\,$K gives rise to a heterogeneous state with rhombohedral crystallites that are embedded in a tetragonal matrix. The nature of the third PT reported at $T < 30\,$K \cite{Lyt64} is less well understood. Measurements with nuclear magnetic resonance (NMR) revealed changes in the local electric field gradients of the Ti-ions that are spatially inhomogenous. The effect is sample dependent and is even strongly affected by oxygen isotope substitution \cite{Bli05, Ito99}, or by the application of lateral stress \cite{Uwe76}.

The cubic-to-tetragonal PT at $T_\mathrm{STO}^{I}$ has been investigated in great detail \cite{Neu95, Wan98, Hir95, Vla00, Mis00, Sal06, Hol07}. For the near-surface region it has been reported that this PT can occur at significanlty higher temperatures of ${T^{I}}'_\mathrm{STO} \approx 150$\,K \cite{Hir95, Vla00, Mis00, Sal06, Hol07}. Furthermore, X-ray diffraction measurements suggest that the near-surface region at ${T^{I}}'_\mathrm{STO} > T > T_\mathrm{STO}^{I}$ consists of a heterogenious mixture of cubic and tetragonal crystallites \cite{Hol07}. This observation highlights that the surface structural properties of  SrTiO$_3$ substrates are subject to complex relaxation phenomena and related structural domain states. It also suggests that an even more complicated behavior can occur if epitaxial layers are grown on top of the STO surface. Significant mutual strain-induced effects between the thin film and the STO near-surface region can be expected here. In an extreme case, the substrate-induced stress may even induce structural phase transitions in the epitaxial layers \cite{Vla00}.

These considerations have motivated us to perform a synchrotron x-ray study of the structural behaviour of Y$_{0.6}$Pr$_{0.4}$Ba$_{2}$Cu$_{3}$O$_{7}$ / La$_{2/3}$Ca$_{1/3}$MnO$_{3}$ superlattices grown epitaxially by pulsed laser deposition on 0.5\,mm thick STO (001) substrates. The superlattices consisted of 10 doublelayers with 10\,nm / 10\,nm layer thicknesses and were grown on untreated substrates at 730$^\circ$C. We used a laser fluence of 1.8\,J/cm$^2$ and an oxygen partial pressure of 0.5\,mbar which resulted in a growth rate of 0.115\,unitcell/s for Y$_{0.6}$Pr$_{0.4}$Ba$_{2}$Cu$_{3}$O$_{7}$ and 0.056\,unitcell/s for La$_{2/3}$Ca$_{1/3}$MnO$_{3}$. The sample was annealed after growth for one hour at 530$^\circ$C in 1.0\,bar oxygen partial pressure~\cite{Hab01}. The measurements were performed at the Material Science (MS) beamline of the Swiss Light Source (SLS) at the Paul Scherrer Institut (PSI) in Villigen, Switzerland. The energy of the x-rays was set to 8.5\,keV with a beam cross section of $2 \times 2\,$mm$^2$. The samples were mounted in a closed cycle refrigerator which covered the temperature range of 18-300\,K. The alignment and rotation of the sample and the detector were achieved with a 2+3-circle surface diffractometer from Micro-Controle Newport equipped with a Physik Instrumente hexapod \cite{Wil05}. The information about the near-surface region of the STO substrate (with an x-ray penetration depth of about 7\,$\mu$m) was obtained from measurements on the STO (002) Bragg reflex (BR) whereas the structural properties of the superlattice have been investigated at grazing incidence angles at the position of the first multilayer Bragg reflex (MBR) (200\,\AA{ }periodicity). To determine the stress and relaxation at the interfaces between STO and the superlattice we mapped the regions near the (103) and (013) BR of STO.

To obtain information about the alignment of the crystallites we performed rocking scans at two sample positions before and after rotating the sample around its surface normal by 90$^\circ$. We refer in the following to these directions as the $a$- and $b$-directions. Due to a slight misalignment of the sample rotation axis and the incident x-ray beam the position of the probed area on the sample surface varies slightly between the measurements for the two orientations. However, we are confident that this is not primarily responsible for the strong anisotropy seen in the data as presented below, as we have observed it reproducably for several samples in different experiments.

      \begin{figure}[!] \center
       \begin{picture}(86,174)
          \put(0.5,-4){\includegraphics[width=85.5\unitlength]{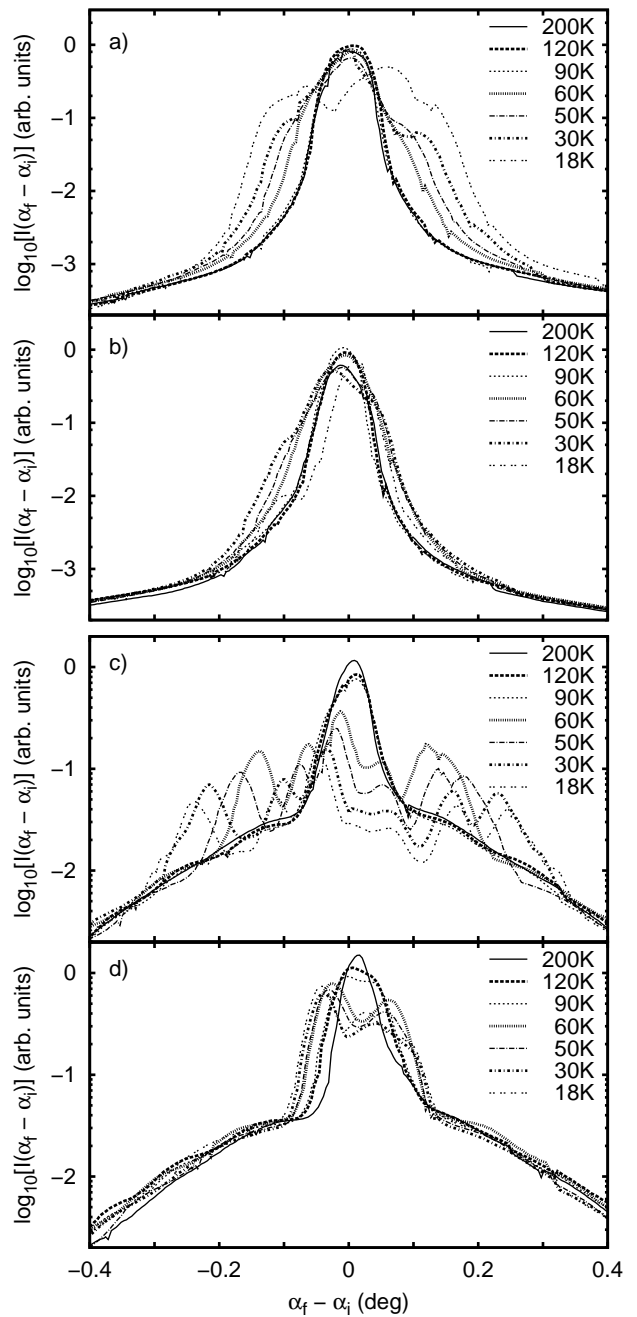}}
        \end{picture}
        \caption{ \label{rocking}
($a$) and ($b$): Rocking curves (plotted as angle of exiting beam $\alpha_\mathrm{f}$ - angle of incident beam $\alpha_\mathrm{i}$) on the SrTiO$_3$ (002) reflex measured along the $a$- and $b$-direction in temperature dependence. ($c$) and ($d$): Rocking curves at the position of the first multilayer Bragg reflex measured along the $a$- and $b$-direction as a function of temperature.
        }
      \end{figure}

Figures~\ref{rocking}~($a$) and ($b$) display rocking curves at the position of the STO (002) BR along the $a$- and $b$-directions, respectively. The lineshapes can be seen to exhibit a distinct broadening below 90\,K which is considerably more pronounced for the $a$- than for the $b$-direction. It consists of a superposition of several BR which are situated at different positions of the rocking curve. This indicates that the x-ray beam probes several crystallites with different $c$-axis orientations. The size of these crystallites must exceed the size of the coherence volume of the incident radiation ($\approx 1\,\mu$m$^3$) but be significantly smaller than the entire probed volume ($\approx 1\,$mm$^3$). Notably, the lineshape broadening is entirely absent at 90\,K while it is already clearly visible at 60\,K. This suggests that the domain formation is related to the structural PT at $T_\mathrm{STO}^{II} \approx 65\,$K rather than to the much discussed antiferrodistortive cubic-to-tetragonal PT at $T_\mathrm{STO}^{I} = 104\,$K. We note that the domain formation which we observe at $T < T_\mathrm{STO}^{II}$ involves a much larger $c$-axis tilting than the one reported in Ref. \cite{Hol07} at $T_\mathrm{STO}^{I} = 104\,$K. Their experiment was sensitive to much weaker distortions since the probed volume was about four orders of magnitude smaller and hence the signal consisted of a superposition of less crystallites with different $c$-axis alignements. Therefore they could observe smaller $c$-axis tiltings.

A second anomaly in the lineshape broadening is observed between 30\,K and 18\,K. Here a narrowing occurs for the $b$-direction as opposed to some additional broadening for the $a$-direction. This unusual behavior occurs in the temperature range where the ordering of the Ti-ions in $^{18}$O-substituted STO in combination with ferroelectricity as well as stress-induced ferroelectricity in ordinary STO have been reported  \cite{Uwe76, Bli05, Ito99}. 

The corresponding structural changes of the Y$_{0.6}$Pr$_{0.4}$Ba$_{2}$Cu$_{3}$O$_{7}$ / La$_{2/3}$Ca$_{1/3}$MnO$_{3}$ superlattice on top of the same STO substrate can be deduced from figures~\ref{rocking}~($c$) and ($d$) which show the temperature dependence of the rocking curves at the position of the first MBR. In the first place, it appears that they undergo a similar temperature dependent broadening as the STO (002) BR. The largest changes occur below $T_\mathrm{STO}^{II}$ where the signal splits up into several MBR (Fig.~\ref{rocking} ($c$)). This splitting indicates the incoherent superposition of reflections originating from surfaces that are tilted relatively to each other. The size of these surface facets  must exceed the lateral coherence volume of the x-ray beam which is of the order of micrometers. Corresponding micrometer-sized structural domains were indeed perviously observed by magneto-optical imaging for La$_{2/3}$Ca$_{1/3}$MnO$_{3}$ on STO \cite{Vla00} which yielded a typical domain size of 10-40\,$\mu$m times several 100\,$\mu$m. From polarised neutron reflectometry measurements on our present superlattices we obtained a similar lower limit of the facet size \cite{Sta05, Hop08}. We observed here a similar splitting of the MBR in the $a$-direction as in the present x-ray study. From the known lateral coherence volume of the neutron beam we were able to derive a lower limit of the facet size of about 20-30\,$\mu$m. Our combined x-ray and neutron data thus both provide clear evidence for strongly anisotropic micrometer-sized facets. The difference in the number of incoherently superposed MBR in Fig.~\ref{rocking}~($c$) and ($d$) suggests that the extent of the facets is three times longer for the $b$-direction than for the $a$-direction. From the total width of the lineshapes we derive that these facets are tilted relative to one another by up to 0.5$^\circ$ along the $a$-direction and 0.2$^\circ$ along the $b$-direction. It appears that this facet pattern involves the entire superlattice including the surface layer of the STO substrate. It is likely caused by the structural PT of the STO substrate as is suggested by the similarities between the rocking scans on the STO (002) BR and the first MBR. Nevertheless, there are also some noticeable differences: For example, the onset of a broadening of the rocking curves on the first MBR occurs already at 120\,K along the $b$-direction (Fig.~\ref{rocking}~($d$)). This suggests that the corresponding surface facets are stabilized by the slight miscut angle of the surface normal with respect to the STO $c$-axis of about 0.26$^\circ$ orientated along the $a$-direction and the subsequent strongly anisotropic terraces with a terrace width of about 86\,nm on the STO surface. The second remarkable difference concerns the changes near $T_\mathrm{STO}^{III}$. A clear anomaly is observed for the STO (002) BR while no corresponding changes are seen at the MBR. 

The overall  behavior suggests that as a function of decreasing temperature, structural domains first develop in the immediate vicinity of the STO substrate surface in the form of phase separated tetragonal and cubic crystallites from where they are propagated into the superlattice (forming the surface facets) but not deeper into the STO substrate (as predominantly probed at the STO (002) BR). Even below the bulk cubic-to-tetragonal PT, our data suggest that these domains (facets) are limited to the vicinity of the STO surface region. This situation suddenly changes at $T_\mathrm{STO}^{II}$, where a fairly ordered pattern of strongly anisotropic crystallites (probably due to embedded rhombohedral crystallites \cite{Bli05}) suddenly develops even several micrometer down into the STO substrate. Curiously, the observed changes below 30\,K seem to be entirely absent in the superlattice and are thus most likely also absent in the topmost surface region of the STO substrate. Since it is well known that $T_\mathrm{STO}^{III}$ is extremely sensitive to small perturbations like $^{18}$O-substitution in STO, it is conceivable that this PT is suppressed due to the strain which the superlattice imposes on the STO surface region. 

      \begin{figure} \center
        \begin{picture}(86,35)
          \put(0,0){\includegraphics[width=89.5\unitlength]{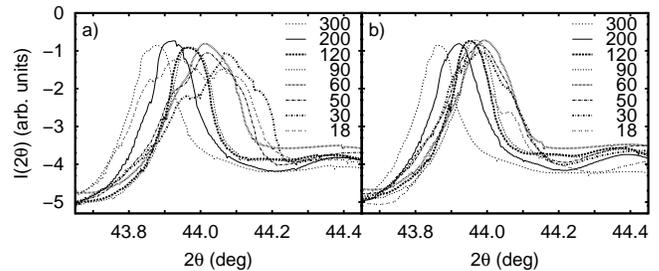}}
        \end{picture}
        \caption{ \label{theta2theta}
Temperature dependent $\theta$ / 2$\theta$ scans at the position of the STO (002) Bragg reflex performed along the crystallographic $a$ and $b$-directions, respectively, probing different volumes near the sample surface.
        }
      \end{figure}

In order to determine the temperature dependence of the $c$-axis lattice parameter of the STO substrate, we also performed $\theta / 2\theta$-scans at the position of the STO (002) BR for the $a$- and $b$-directions and hence the respectively probed volumes as shown in (Figure~\ref{theta2theta}). In the cubic state we obtain $c$-axis lattice parameters of 0.3905, 0.3898 and 0.3896\,nm at 300, 120, and 90\,K, which agree well with the tabulated values in Ref.~\cite{Lyt64}. The 60\,K-scans reveal crystallites with different $c$-axis lengths between 0.3887\,nm and 0.3899\,nm for the volume probed in the scans in direction $a$ and a somewhat smaller spread from 0.3893\,nm to 0.3899\,nm for the volume probed along direction $b$. At 30\,K, the $c$-axis lengths get even shorter in the volume along the $a$-direction while it remained almost the same in the one probed along $b$. Below 30\,K, another PT takes place, as already shown in the rocking scans in figure~\ref{rocking}~($a$) and ($b$). At 18\,K, the $c$-axis lengths are between 0.3884\,nm and 0.3913\,nm in the volume probed along the $a$-direction, while they only vary between 0.3889\,nm and 0.3906\,nm for the $b$-direction. A similar sudden increase in the $c$-length at low temperatures has also been reported in ref. \cite{Lyt64}.  

      \begin{figure} \center
        \begin{picture}(86,57)
          \put(-3,-3){\includegraphics[width=90\unitlength]{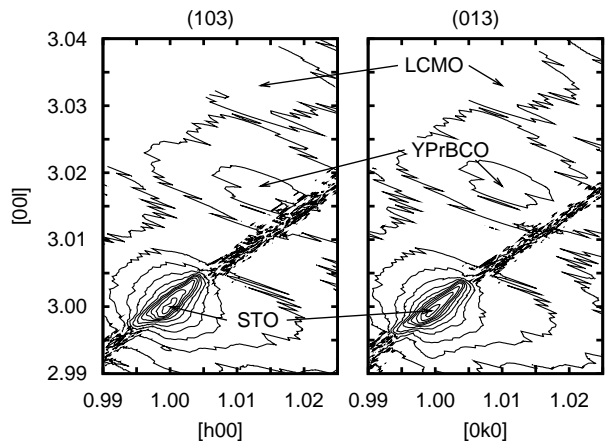}}
        \end{picture}
        \caption{ \label{asymmetric}
Mappings of the asymmetric  (103) and (013) STO Bragg reflexes at 200\,K. The Y$_{0.6}$Pr$_{0.4}$Ba$_2$Cu$_3$O$_7$ (109) and (019) Bragg reflexes are visible at (1.014 0 3.18) and (0 1.01 3.18), respectively while the La$_{2/3}$Ca$_{1/3}$MnO$_{3}$ (161) Bragg reflex is visible at (1.014 0 3.033) and (0 1.01 3.033). The diagonal lense shape of the STO Bragg reflexes results from the instrument resolution (detector streak), while the diagonal line of noisy signal is due to the filters used to protect the 2D-detector from oversaturation at the position of the main STO peak. The linespacing of the contours corresponds to $0.5 \times \log_{10}[I_\mathrm{measured}]$.
        }
      \end{figure}

To determine the strain and the relaxation at the interface between the STO substrate and the superlattice, we mapped the regions near the (103) and (013) asymmetric BR of STO.  Figure~\ref{asymmetric} shows the mappings that were obtained at 200\,K by performing rocking scans at the respective positions and measuring the diffracted intensities with an area detector. The crystal structure of STO could be determined from the main peaks as being cubic with a lattice parameter of 0.39005\,nm. We note that the lens-like shape of the peaks along the diagonal of the maps is caused by the detector streak and the instrument resolution. The noisy signal on the diagonal line through the main peak is caused by filters inserted in the direct beam to avoid oversaturation in the 2D detector and hence reducing the statistics at the measured point. The shoulders of the main peaks towards larger $h$ and $k$ values but smaller $l$ values indicate that the STO unit cells nearest to the multilayer structure exhibit a lateral shrinking combined with a slightly increased $c$-axis parameter to fit the Y$_{0.6}$Pr$_{0.4}$Ba$_2$Cu$_3$O$_7$ $ab$-plane. The obtained lattice parameters of Y$_{0.6}$Pr$_{0.4}$Ba$_2$Cu$_3$O$_7$ are $a = 0.385\,$nm, $b = 0.386$\,nm and $c = 1.163\,$nm which are in good agreement with the tabulated values of $a = 0.38334\,$nm, $b = 0.39034\,$nm and $c = 1.1686\,$nm at 300\,K \cite{Ber96}. Its (109) and (019) BR are located at (1.014 0 3.018) and (0 1.01 3.018), respectively. La$_{2/3}$Ca$_{1/3}$MnO$_{3}$ is found to be unrelaxed with the lattice parameters $a = 0.546\,$nm, $b = 0.547$\,nm and $c = 0.711\,$nm, as given in \cite{Vla00, Bla96}, where the larger, orthorhombic La$_{2/3}$Ca$_{1/3}$MnO$_{3}$ unit cell is rotated by 45$^\circ$ around the $c$-axis as compared to the one of Y$_{0.6}$Pr$_{0.4}$Ba$_2$Cu$_3$O$_7$. Its (161) BR is located at (1.014 0 3.033) and (0 1.01 3.033).

We conclude from our observations in combination with the ones of earlier reports that the structure of the near-surface region of STO is spatially inhomogeneous. The antiferrodistortive PT begins already at ${T^{I}}'_\mathrm{STO} \approx 150$\,K and yields a mixture of tetragonal and cubic crystallites. Around $T_\mathrm{STO}^{II} = 65\,$K, a sizeable volume fraction of rhombohedral crystallites form in the otherwise tetragonal matrix. This leads to anisotropic surface facets that are tilted relatively to each other by up to 0.5$^\circ$ along the direction of the miscut of the substrate surface, while the tilting along the other direction remains considerably smaller. The multilayers grown epitaxially on top of these substrates follow the tilting but may force a relaxation of the uppermost STO-layers. Our observations have important implications, especially in the context of the recent attempts to obtain novel two-dimensional electronic and magnetic states via interface engineering of oxide-based multilayers. Our study shows, that the strain effects induced by the STO substrate can be sizeable and most importantly can develop at substantially different temperatures than the well known antiferrodistortive PT of bulk STO.

We acknowledge R.F.~Kiefl for the fruitful discussions and D.~Meister for the technical assistance. The work was performed at the Swiss Light Source at Paul Scherrer Institut and has been supported by the Swiss National Foundation with grant 200020-119784-1 and the NCCR Materials with Novel Electronic Properties - MaNEP.

\end{document}